\documentclass[DIV=calc,paper=a4,fontsize=11pt,twocolumn]{scrartcl} 

\usepackage[english]{babel}
\usepackage[protrusion=true,expansion=true]{microtype}
\usepackage{amsmath,amsfonts,amsthm}
\usepackage[final]{graphicx}
\usepackage{xcolor}
\usepackage[normal,small,hypcap,up,labelfont=bf,textfont=it]{caption}
\usepackage{epstopdf}
\usepackage{subfig}
\usepackage{booktabs}
\usepackage{fix-cm}
\usepackage{amssymb,amsfonts}
\usepackage{dsfont}
\usepackage{bbm}
\usepackage{pstricks}
\usepackage{cite}
\usepackage[utf8]{inputenc}
\usepackage[perpage,symbol*]{footmisc}
\usepackage[varg]{txfonts}
\usepackage{balance}
\usepackage{fancyhdr}
\usepackage{bookmark}
\usepackage{verbatim}
\usepackage{fontenc}
\usepackage{cuted}

\usepackage{bm}
\usepackage{mathrsfs}

\theoremstyle{definition}

\theoremstyle{plain}

\DeclareCaptionFont{mycolor}{\color[HTML]{0000FF}}
\usepackage{sectsty}													
\allsectionsfont{
\color[HTML]{31ADF3}\usefont{OT1}{phv}{b}{n}
}

\sectionfont{
\color[HTML]{31ADF3}\usefont{OT1}{phv}{b}{n}
}

\usepackage{fancyhdr}												
\usepackage{lettrine}
\newcommand{\initial}[1]{%
\lettrine[lines=3,lhang=0.3,nindent=0em]{
\color[HTML]{31ADF3}
{\textsf{#1}}}{}}

\usepackage{titling}															

\newcommand{\HorRule}{\color[HTML]{31ADF3}
\rule{\linewidth}{1pt}%
}

\pretitle{\vspace{-30pt} \begin{flushleft} \HorRule
\fontsize{34}{34} \usefont{OT1}{phv}{b}{n} \color[HTML]{31ADF3} \selectfont
}

\title{Accommodating Retrocausality with Free Will}					
\posttitle{\par\end{flushleft}\vskip 0.5em}

\preauthor{\begin{flushleft}\large \lineskip 0.5em \usefont{OT1}{phv}{b}{sl} \color[HTML]{31ADF3}}
\author{Yakir Aharonov$^{~\mathsf{1,2}}$, Eliahu Cohen$^{~\mathsf{3,1}}$ \& Tomer Shushi$^{~\mathsf{4}}$\\[8pt]}											
\postauthor{\footnotesize \usefont{OT1}{phv}{m}{sl} \color[HTML]{000000}
$^{\mathsf{1}}$ School of Physics and Astronomy, Tel Aviv University,\\
Tel-Aviv, Israel. E-mail: \href{mailto:eliahuco@post.tau.ac.il}{eliahuco@post@post.tau.ac.il}\\
$^{\mathsf{2}}$ Schmid College of Science, Chapman University, Orange,\\
California, USA. E-mail: \href{mailto:yakir@post.tau.ac.il}{yakir@post.tau.ac.il}\\
$^{\mathsf{3}}$ H.H. Wills Physics Laboratory, University of Bristol, Tyndall Avenue,\\
Bristol, UK. E-mail: \href{mailto:eliahu.cohen@bristol.ac.uk}{eliahu.cohen@bristol.ac.uk}\\
$^{\mathsf{4}}$ University of Haifa, Haifa 27019, Israel. E-mail: \href{mailto:tomer.shushi@gmail.com }{tomer.shushi@gmail.com }\\[10pt]		
\scriptsize\usefont{OT1}{phv}{m}{n} \color[HTML]{31ADF3}{\textbf{Editors: \emph{First Editor}, \emph{Second Editor} \& \emph{Third Editor}} }\\[5pt]
\color[HTML]{000000}{Article history: Submitted on Month Day, 2014;  Accepted on Month Day, 2014; Published on Month Day, 2014.}
\par\end{flushleft}\HorRule}

\date{}																				

\begin{document}
\maketitle
\thispagestyle{fancy} 			
\initial{R}\textbf{etrocausal models of QM add further weight to the
conflict between causality and the possible existence of free will. We
analyze a simple closed causal loop ensuing from the interaction between two
systems with opposing thermodynamic time arrows, such that each system can
forecast \textquotedblleft future\textquotedblright\ events for the other.
The loop is avoided by the fact that the choice to abort an event thus
forecasted leads to the destruction of the forecaster's past. Physical law
therefore enables prophecy of future events only as long as this prophecy is
not revealed to a free agent who can otherwise render it false. This
resolution is demonstrated on an earlier finding derived from the
Two-State-Vector Formalism (TSVF), where a weak measurement's outcome
anticipates a future choice, yet this anticipation becomes apparent only
after the choice has been actually made. To quantify this assertion,
\textquotedblleft weak information\textquotedblright\ is described in terms
of Fisher information. We conclude that an \textquotedblleft already
existing\textquotedblright\ future does not exclude free will nor invoke
causal paradoxes. On the quantum level, particles can be thought of as
weakly interacting according to their past and future states, but causality
remains intact as long as the future is masked by quantum indeterminism.%
\newline
Quanta 2015; 4: xx--yy.}

\begin{figure}[b!]
\rule{245 pt}{0.5 pt}\\[3pt]
\raisebox{-0.2\height}{\includegraphics[width=5mm]{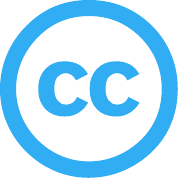}}\raisebox{-0.2%
\height}{\includegraphics[width=5mm]{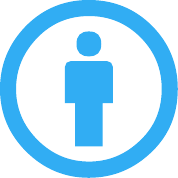}} {\footnotesize {This is an open
access article distributed under the terms of the Creative Commons
Attribution License \href{http://creativecommons.org/licenses/by/3.0/}{%
CC-BY-3.0}, which permits unrestricted use, distribution, and reproduction
in any medium, provided the original author and source are credited.} }
\end{figure}

\section{Introduction}

\label{Intro}

Time-symmetric formulations of QM are gaining growing interest. Using two
boundary conditions rather than the customary one, they offer novel twists
to several foundational issues. Such are the Wheeler-Feynman electromagnetic
absorber theory \cite{Wheeler}, and Hoyle and Narlikar's modification \cite%
{Hoyle} and Cramer's transactional interpretation \cite{Cramer}. Among
these, however, the ABL rule \cite{Y1} and Aharonov's Two-State-Vector
Formalism (TSVF) \cite{Y2} are distinct, in that they even predict some
novel effects for a combination of forwards and backwards evolving wave
functions. When performing a complete post-selection of the quantum state,
otherwise counterfactual questions can be intriguingly answered with regard
to the state's previous time evolution.

These advances, however, might seem come with a price that even for adherents
is too heavy, namely, dismissing free will. While quantum indeterminism
seemed to offer some liberation from the chains imposed on our choices by
classical causality, time-symmetric QM somewhat undermines quantum
indeterminism, as it renders future boundary conditions the missing source
of possible causes. This might eventually reveal causality to be just as
strict and closed as classical causality. If the future is, in some sense,
\textquotedblleft already there\textquotedblright\ to the point of being
causally equal to the past, free will (which is defined in the next section)
might appear to be as illusory as it has appeared within the classical
framework. We aim to show this is not necessarily the case. TSVF is no worse
off than classical physics, or other formulations of quantum mechanics as it pertains to
the incorporation of free will. In other words, free will is not precluded
even when discussing a quantum world having both past and future boundary.

In this special issue of Quanta, dedicated to Richard Feynman and discussing
time-symmetry in quantum mechanics, we examine what might seem to be a problem
in these formulations, namely the notion of free will \cite{Russel1}.
Discussion of this kind might at first be regarded as philosophical in
character, but we hope to formulate the problem rigorously enough to yield
nontrivial physical insights.

\section{The Problem}

Following Russell and Deery \cite{Russel1}, we propose defining free will as
follows. Let a physical system be capable of initiating complex interactions
with its environment, gaining information about it and predicting its future
states, as well as their effects on the system itself. This grants the
system purposeful behavior, which nevertheless fully accords with \textit{%
classical} causality. Now let there be more than one course of action that
the system can take in response to a certain event, which in turn lead to
different future outcomes that the system can predict. \textquotedblleft
Free will\textquotedblright\ then denotes the system's taking one out of
various courses of action, independently (at least to some extent) of past
restrictions. This definition is very close in spirit to the one employed
in \cite{Georgiev}, i.e. the ability to make choices. It should be
emphasized that even in our time-symmetric context, free will means only
freedom from the past, not from the future (see also \cite{Y2}).

In classical physics, conservation laws oblige any event to
be strictly determined by earlier causes. In our context, this might apparently leave only
one course of action for the system in question, and hence no real choices. When moving to the quantum realm, free will might be recovered \cite{Georgiev}, but then again, if one adds a final boundary condition to the description of the quantum system, can free will exist? We shall answer both classical and quantum questions on the affirmative, employing statistical and quantum fluctuations, respectively.

In what follows, we analyze a classical causal paradox avoided by the past's
instability. We subsequently consider a more acute variant of this paradox and discuss a few possible resolutions. Then we present the quantum counterpart of
these two paradoxes where inherent indeterminism saves causality. We show that within
the TSVF, although both future and past states of the system are known,
genuine freedom is not necessarily excluded. We
then define and quantify the term \textquotedblleft weak
information\textquotedblright\ -- the kind of information coming from the
future that can be encrypted in the past without violating causality.

\section{An Interaction between Two Systems with Opposing Time-Arrows}

To demonstrate the possibility of knowing one's future and its consequences,
we discuss a highly simplified classical gedanken experiment. Naturally,
there are immediate difficulties with such a setup. Can, e.g., two regions
in space have opposite time arrows to begin with? Can observers inside them
communicate? etc. These questions deserve further probing, but we focus here
only on what would happen if several conditions are met, rather than whether
and how they can be achieved.

Consider, then, a universe comprised of only two closed, non-interacting
laboratories located at some distance from one another. Suppose further that
their thermodynamic time arrows are opposite to one another, such that each
system's \textquotedblleft future\textquotedblright\ time direction is the
other's \textquotedblleft past\textquotedblright .\ Finally let each
laboratory host a free agent, henceforth Alice and Bob, capable of free
choice.

It is challenging to create a communication channel between two laboratories
of this kind. An exchange of signals is possible in the following form. A
light beam is sent from the exterior part of one laboratory to the other's
boundary, where a static message is posted. The beam is then reflected back
to its origin. If the labs are massive enough, the beam imparts only a
negligible momentum transfer.

\begin{figure}[t!]
\centering
\includegraphics[width=60mm]{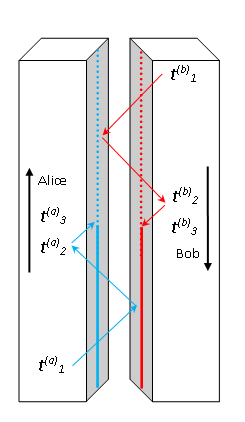}
\caption{\color[HTML]{0000FF}{\label{fig:1} An illustration of the two
labs
gedanken experiment with free agents.}}
\end{figure}

The gedanken experiment is as follows (see Fig. \ref{fig:1}):

$t^{(b)}_1$: Bob sends a light-beam (red arrow) to Alice's lab.

$t^{(b)}_2$: He receives through his returning beam a message from Alice
saying: \textquotedblleft Let me know if you see this
message\textquotedblright\ (dotted blue world-line).

$t^{(b)}_3$: Bob posts a confirmation saying: \textquotedblleft I saw your
message\textquotedblright (red world-line).

Then there are the following events in Alice's lab:

$t_{1}^{(a)}$: Alice sends a light-beam (blue arrow) to Bob's lab.

$t_{2}^{(a)}$: Alice receives through her returning beam of particles that
scattered off Bob's message, i.e. she gets the information from Bob through
this beam reflected from Bob's system to her system.

$t^{(a)}_3$: Alice, realizing that this confirmation comes from her \textit{%
future}, chooses not to post a message.

The Causal Paradox is obvious: The dotted blue world-line represents an
absent message. How, then, could Bob reply to a message which was removed
before he was supposed to see it?

It should be noted, that alongside with this formulation of the paradox, one
can equivalently describe the complementary scenario: Bob finds through his
returning beam that Alice did not post a message. Therefore, he sends no
confirmation, but eventually Alice, having free will, decides to post a
message in contrast to Bob's observation.


\section{The Suggested Resolution}

A key element in this causal paradox's resolution is the following
well-known fact: Entropy-increasing processes are highly stable, not
sensitive to small changes in their initial conditions or their evolution,
whereas entropy-decreasing processes are extremely vulnerable to any
interference.

Our question therefore is: Which time direction is affected by Alice's
decision to change the \textquotedblleft future\textquotedblright that
\textquotedblleft has been forecasted\textquotedblright\ by Bob? The
simplest and most consistent answer is: Bob's past. Upon Alice's decision to
remove her message at t(a)3, Bob's \textquotedblleft
prophecy\textquotedblright, i.e. the message of Bob to Alice regarding her
future choice, turns out to be false. This is clearly inconsistent with his
earlier observation of Alice's message, which is understood now to be highly
unstable. His observation turns out to be a large (hence very rare)
statistical fluctuation.

We can now define the arrow of time of any system as the thermodynamic
direction which is stable against changes. While a small change at the large
system's present will negligibly affect its future, it can have dramatic
effects into its past. Alice's future was coupled in our example to Bob's
past. By employing her free will, she could completely alter his previous
observations, but the apparent paradox is resolved by taking into account
the chaotic nature of the entropy decreasing direction. Indeed, the signals
are weak enough, which makes them amenable for this reinterpretation as
fluctuations.

\section{A More Acute Paradox}

\begin{figure}[t!]
\centering
\includegraphics[width=84mm]{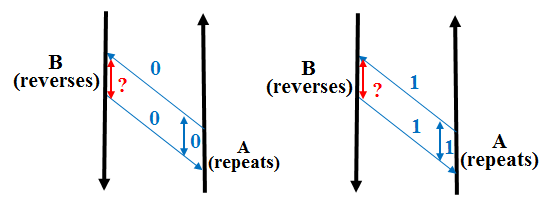}
\caption{\color[HTML]{0000FF}{\label{fig:2} An illustration of
the
two
machines gedanken experiment. The paradox is symmetric, but
for
simplicity
it is shown to reside on the B side. }}
\end{figure}

We shall now discuss an operationally simpler, yet conceptually harder
version of the paradox, which emphasizes the role of free will. Let the two
labs with opposing time arrows contain two simple machines rather than free
agents (see Fig. \ref{fig:2}). One machine, A, posts 0 if it receives 0 as
an input, and 1 if it receives 1. The second lab's machine, B, posts 1 if it
receives 0 and 0 if it receives 1.\newline
The paradox is as follows: In case A receives 0 from the other lab, it posts
0. Then B receives the 0 as an input and posts 1, in contrast to A's earlier
input. Alternatively, A receives 1 from B, then posts 1. Then, B receives
this 1 and posts 0, again in contrast to the A's initial input. \newline
It follows there are no valid initial conditions for this combined system at
a given time. \newline
The resolution may be: \newline
(1) Communication is impossible between two such systems. \newline
(2) The past of both systems is symmetrically unstable. \newline
(3) There must be some stochastic element allowing consistency.\newline
(4) The operations of the two machines must be coordinated. \newline
As explained above, we assume that communication of simple static messages
is possible, hence we shall avoid the first option (nevertheless, this paradox could actually suggest that a special communication protocol is needed between
two such systems with opposite time arrows). Options (2) and (3) complement
each other and resonate with the above notion of free will, as well as with
the quantum paradox to be presented below. Naturally, this combination is
favored by us. We believe this paradoxical situation could have been avoided
if a minor degree of freedom (e.g. at the form of free will) were allowed.
In contrast, alternative (4) implies superdeterminism (see for instance \cite{'tHooft1,'tHooft2}) or the so called
``conspiracy'' between the two machines, which is philosophically disturbing
(at least in our view), negating free will altogether.

\section{Going Quantum: The TSVF and Weak Measurements}

The possibility for resolving the above problem on classical grounds
encourages seeking more interesting avenues at the quantum level. Indeed a
similar resolution will be offered, i.e. the possibility of re-interpreting
the past. However, the basic concept on which the resolution relies shifts
from thermodynamic to quantum fluctuations which are more suitable for
describing small microscopic systems. This is where time-symmetric quantum
causality comes in most naturally.

TSVF is a time-symmetric formulation of quantum mechanics employing in
addition to the forward evolving wave function (pre-selected state) also a
backwards evolving wave function (post-selected state). This combination
gives rise to the \textit{two-state-vector}, which provides a richer notion
of quantum reality between two projective measurements. This world-view has
produced several predictions, so far well verified by weak measurements \cite%
{Y3,TC,Y4,Y5} which delicately gather information about the quantum state
without collapsing it, and thus do not change to post-selection probability.

In an earlier work \cite{Y5} the following gedanken experiment was proposed.
A large ensemble of $N$ EPR spins is prepared. Each particle in every pair
is weakly measured along the three Bell orientations, before being strongly
projected along one of them. As was shown, each weak measurement only
slightly disturbs the state and hence the well-known non-local correlations
between the strong outcomes are maintained in this experiment. It should be
noted that in return each weak measurement provides only a negligible amount
of information (to be quantified in the next Section). However, since all
the weak outcomes were classically recorded, upon slicing them according to
the projective outcomes, one finds in retrospect, with extreme accuracy, the
weak values corresponding to all Bell orientations (not only the ones
eventually chosen for the projective measurement). The question is then, how
could the values reside in the weak data prior to the final Bell
measurements which demonstrated almost perfect non-local correlations?
Bell's proof certainly forbids them to be prepared in such a way so the TSVF
answer would be they came from the future! The important point in this
retrocausal interpretation is the weak values could be there, that is, could
had causal effect on the pointer's shift, without forcing a specific future
outcome.

The resolution is therefore simple: Quantum indeterminacy guarantees that,
should someone try to abort a future event about which they have received a
prophecy, that prophecy would turn out to be a mere error.

Therefore, even in the TSVF where present is determined by both past and
future events, the quantum indeterminism enables free will.

Naturally, more mundane explanations ought to be considered before
concluding that results of weak measurement contain information regarding a
future event. By normal causality, it should be Alice's measurements which
affected Bob's, rather than vice versa. Perhaps, for example, some subtle
bias induced by her weak measurements affected his later strong ones.

Such a \textquotedblleft past-to-future\textquotedblright\ effect is
considerably strained by the following question: How robust is the alleged
bias introduced by the weak measurements? If it is robust enough to oblige
the strong measurements, then it is equivalent to full collapse, namely the
very local hidden variables already ruled out by Bell's inequality. This is
clearly not the case: weakly measured particles remain nearly fully
entangled. But then, even the weakest bias, as long as it is expected to
show up over a sufficiently large N, is ruled out of the same grounds. The
\textquotedblleft weak bias\textquotedblright\ alternative is ruled out also
by the robust correlations predicted between all same-spin measurements,
whether weak or strong.

Can Alice predict Bob's outcomes on the basis of her own data? To do that,
she must feed all her rows of outcomes into a computer that searches for a
possible series of spin-orientation choices plus measurement outcomes, such
that, when she slices her rows accordingly, she will get the complex pattern
of correlations described above.

The number of such possible sequences that she gets from her computation is $%
\left(
\begin{array}{c}
N \\
N/2%
\end{array}%
\right) \propto \frac{2^{N}}{\sqrt{N}}$. Each such sequence enables her to
slice each of her rows into two $N/2$ halves and get the above correlations
between her weak measurements and the predicted strong measurements. Notice
that, according to Sec. 5, the results' distribution is a Gaussian with $%
\lambda \sqrt{N}/2$ expectation and $\delta \sqrt{N}/2$ standard deviation,
so a $\delta $ shift in one of the results, or even in $\sqrt{N}$\ of them,
is very probable. Hence, even if Alice computes all Bob's possible future
choices, she still cannot tell which choice he will take, because there are
many similar subsets giving roughly the same value. Also, as Aharonov et al.
pointed out in \cite{Y5}, when Alice finds a subset with a significant
deviation from the expected 50\%-50\% distribution, its origin is much more
likely, upon a real measurement by Bob, to turn out to be a measurement
error than a genuine physical value. Obviously, then, present data is
insufficient to predict the future choice.

\section{The \textquotedblleft strength\textquotedblright\ of Information
Transmission}

The information transmission between Alice and Bob can be categorized into
two different types with different \textquotedblleft
strength\textquotedblright :

1. \textit{Strong information}: This type describes the information that, in
general, has the potential to interfere with Alice's free choice. This is
the classical kind of information transmitted in the first gedanken
experiment.

2.\textit{Weak information}: In this case the information that Bob sends to
Alice will not, in any circumstance, interfere with Alice's free choice
because it is buried much below the quantum uncertainty level.

While the strong information transmission was discussed in Secs. 3-4 and was
shown to cast instability into Bob's past, it seems the weak information
notion should be further explained and quantified.

We now understand that weak information represents information that does not
actively interfere with the Alice's and Bob's systems. Therefore, weak
information can be described employing weak measurement outcomes since
individually they only provide very partial information that does not
interfere neither Alice's nor Bob's system consistency. Similarly, strong
information is related to projective measurement outcomes since they do
disturb the systems and provide definite results.

To create a clear distinction between the two kinds of information, we shall discuss a simple thought experiment. Suppose Alice has a spin she wants to measure. To do that, she will use a Stern-Gerlach magnet with a non-homogeneous magnetic field along some direction. Bob, having an opposite time arrow, already knows that Alice will choose the z-axis and and will find an ``up'' outcome. If Bob sends this (strong) information to Alice, she may choose the y-axis instead and find a ``down'' outcome, reproducing the paradoxical situation discussed above. However, if Bob only tells her she will find an ``up'' result along {\it some} direction, no causal paradox will ensue (see also \cite{ECS}. This is the kind of weak information which does not clash with Alice's free will nor with Bob's history.



\section{Fisher Information for Strong and Weak measurements}

Fisher information is a tool to quantify the hidden information in a random
variable $Q$ regarding a parameter it depends on. Using Fisher information we can now quantitatively define the strong and
weak information concepts that were qualitatively introduced in the Section
above.

Suppose there is an unknown parameter $\theta $ which we want to estimate ($%
\theta$ can stand for, e.g., the relative phase between two superposed
wave-packets). We define a density function of $Q$ by $f$, and another auxiliary parameter $\Delta$ which describes the type of information, strong or weak. In probabilistic terms, it is called the \textquotedblleft scale
parameter\textquotedblright\ of $Q$. In this case, Fisher information as a
function of $\Delta $, $I_{\Delta }(\theta )$, is given by%
\begin{equation}
I_{\Delta }(\theta ):=E\left( \left[ \frac{\partial }{\partial \theta }\ln
f\left( \Delta Q;\theta \right) \right] ^{2}|\theta \right) .
\end{equation}%
It can be easily shown that $I_{\Delta }(\theta )$\ is in fact the product
of $\Delta ^{-1}$\ and $I(\theta )$:%
\begin{eqnarray}
I_{\Delta }(\theta ) &=&\int \left[ \frac{\partial }{\partial \theta }\ln
f\left( \Delta Q;\theta \right) \right] ^{2}f\left( \Delta Q;\theta \right)
dQ \\
&=&\frac{1}{\Delta }\int \left[ \frac{\partial }{\partial \theta }\ln
f\left( Q;\theta \right) \right] ^{2}f\left( Q;\theta \right) dQ \\
&=&\Delta ^{-1}I(\theta ).
\end{eqnarray}%
Now, in case of $\Delta \rightarrow 0$, we find%
\begin{equation}
\underset{\Delta \rightarrow 0}{\lim }I_{\Delta }(\theta )=\underset{\Delta
\rightarrow 0}{\lim }\Delta ^{-1}I(\theta )=\infty ,
\end{equation}%
hence we conclude that $\Delta \rightarrow 0$\ indicates strong information.

The opposite case of $\Delta \rightarrow \infty $\ leads to

\begin{equation}
\underset{\Delta \rightarrow \infty }{\lim }I_{\Delta }(\theta )=\underset{%
\Delta \rightarrow \infty }{\lim }\Delta ^{-1}I(\theta )=0,
\end{equation}%
which implies weak information, so for sufficient large value of $\Delta $,
weak information is described by a negligible Fisher information.

Let us now demonstrate this concept. Suppose that $\theta $ is the relative phase between two superposed
wave-packets, which we want to measure in some interference experiment.  Let us assume that the interference pattern is
detected via some coupling $1/\Delta$ to a measuring pointer. If our estimation for the relative phase is described by a Gaussian random variable $Q$, then the
density function of $Q$ will be $f(Q;\theta )=\frac{1}{\sqrt{2\pi }\theta }%
\exp \left( -\frac{1}{2\theta ^{2}}Q^{2}\right) $. Depending on the coupling strength, the Fisher information will be $I_{\Delta }(\theta
)=\Delta ^{-1}\theta ^{-2}.$

\section{Cryptography Can Protect Causality}

Weak or encrypted information can be used for communication between future and past in a causality preserving manner thanks to quantum indeterminism.
The main idea behind this type of communication is quantum cryptography  \cite{Gisin}.  Suppose Bob somehow knows what Alice will choose in the future. He uses a quantum cryptography scheme to encode Alice's future choice and gives her the encrypted prophecy. However, he does not share with her the key to decode this revelation until she actually makes her choice. In this case, similarly to the example in Sec. 6, both Bob's past and Alice's future are secured. Due to quantum indeterminism, Alice still has free will.

For example, in the BB84 scheme \cite{BB84}, even though Alice and Bob communicate through a public channel, their secret key is secured due to another form of quantum indeterminism, namely, that non-orthogonal states are indistinguishable. This means that even if the generated string contains information regarding Eve's future, it will not create a causal paradox.

\section{A Few alternatives}

In addition to the proposed resolution for the above paradoxes, there exist
some other well-known possibilities. The parallel universe resolution
suggests that if one goes back in time and kills his grandfather he actually
does it in a parallel universe and therefore he does not interfere with the
laws of nature \cite{Bousso,Deutch}. A different approach to solve this is
by postulating another time dimension in which such disagreements can be
solved before being recorded in our history \cite{Bars1,Bars2}.

These two resolutions clearly lack simplicity and oblige an excessive
ontology to our existing theories. Moreover, detailed work is needed to
refute each and every paradox.

Therefore, bearing in mind Occam's razor as a tool for denying complex
theories, it seems these alternative solutions are unfavorable.

Another solution simply dictates that one cannot create paradoxes in the
universe and therefore cannot, for instance, kill his grandfather.
This approach implies a universe guided by global consistency condition such
as in \cite{Carlini,Novikov} and was shown to naturally arise in post-selected closed-timelike curves \cite{Lloyd}.

\section{Free will and Becoming}

Classical physics treats time as a purely geometrical ingredient of the
universe, alongside the three spatial dimensions. Against the perfect
logical rigor and experimental support that make relativity so powerful,
many physicists find the \textquotedblleft block universe\textquotedblright\
picture emerging from it manifestly awkward. In fact, the very notion of
space-time implies that, just as all locations have the same degree of
reality in space, so do all past, present, and future moments exist along
the temporal dimension without any moment being unique as the privileged
\textquotedblleft now \textquotedblright.

Against this mainstream view, there are alternative accounts \cite{Y6}. They
suspect that, if we experience time so differently from space, this
difference may be objective. They provide some models to capture this notion
of dynamic time.

Bob's access to Alice's future in the classical gedanken experiment above
and the double boundary condition on the wave function proposed by the TSVF
may seem at first sight to resonate with a block universe approach. However,
as we have just seen, statistical and quantum fluctuations may provide us
with freedom to define the present. As was shown in Secs. 4-5, this freedom,
and also the notion of becoming, is subjective and system-dependent.

Within the TSVF, while both backward and forward states evolve
deterministically, they have limited physical significance on their own ---
physical reality is created by the product of the causal chains extending in
both temporal directions. The past does not determine the future,yet the
future is set, and only together do they form the present. However, the
existence of a future boundary condition,and its deterministic effect, do
not deny our freedom of choice. It is allowed due to the inaccessibility of
the data (which is a requirement of causality, as discussed in Sec. 5).
Examining the concept of free will from a physical point of view,we find it
must contain at least partial freedom from past causal constraints, and such
freedom is duly manifested in the TSVF, where a juxtaposition of freedom and
determinacy is epitomized.

\section{Conclusions}

We examined the possibility of free will in a retrocausal theory. Closed
causal loops, which arise due to the interaction between two systems with
opposing time arrows were discussed. The suggested resolution of the ensuing
paradoxes relies on the thermodynamic instability of the past.

Moving to the quantum realm, a similar paradox can be solved via the quantum
indeterminism, which is understood to protect free will. This resonates with
previous findings of Georgiev \cite{Georgiev}. Furthermore, we discussed the
\textquotedblleft strength\textquotedblright\ of information transmission,
where the terms \textquotedblleft strong\textquotedblright\ and
\textquotedblleft weak\textquotedblright\ are related to strong (projective)
and weak values, respectively. When information about a future event is
buried under quantum indeterminism it cannot violate free will. Similarly,
encrypted information, e.g., the one available through weak measurements,
does not violate causality. The existence of free will in these time
symmetric models was conjectured to resonate with a dynamical notion of time.

\section*{Acknowledgements}

We thank A.C. Elitzur and Nissan Itzhaki for very helpful discussions. Y.A.
and E.C. acknowledge support of the Israel Science Foundation Grant No.
1311/14 and ERC-AD NLST. Y.A. also acknowledges support from ICORE
Excellence Center \textquotedblleft Circle of Light\textquotedblright.


\begin{thebibliography}{99}
\bibitem{Wheeler} Wheeler JA, Feynman RP. \newblock Interaction with the
absorber as the mechanism of radiation. \newblock \emph{Reviews of Modern
Physics} 1945 17(2--3): 157--161. \newblock \href{http://journals.aps.org/rmp/abstract/10.1103/RevModPhys.17.157}%
{\path{doi:10.1103/RevModPhys.17.157}}.

\bibitem{Hoyle} Hoyle F, Narlikar JV. \newblock A new theory of gravitation. %
\newblock \emph{Proceedings of the Royal Society of London-Series A:
Mathematical, Physical and Engineering Sciences} 1964 282(1389): 191--207. %
\newblock\href{http://rspa.royalsocietypublishing.org/content/282/1389/191.short}{%
\path{doi:10.1098/rspa.1964.0227}}.

\bibitem{Cramer} Cramer JG. \newblock The transactional interpretation of
quantum mechanics. \newblock \emph{Reviews of Modern Physics} 1986 58(3):
647--687. \newblock\href{http://journals.aps.org/rmp/abstract/10.1103/RevModPhys.58.647}%
{\path{doi:10.1103/RevModPhys.58.647}}.

\bibitem{Y1} Aharonov Y, Bergmann PG, Lebowitz JL. \newblock  Time symmetry
in the quantum process of measurement. \newblock \emph{Physivsl Review}
1964; 134(6): B1410. \newblock \href{http://dx.doi.org/10.1103/PhysRev.134.B1410}%
{\path{doi:10.1103/PhysRev.134.B1410}}.

\bibitem{Y2} Aharonov Y, Cohen E, Gruss E, Landsberger T. \newblock %
Measurement and collapse within the Two-State-Vector Formalism. \newblock
\emph{Quantum Studies: Mathematics and Foundations} 2014; 1(1-2):133--146. %
\newblock \href{http://link.springer.com/article/10.1007/s40509-014-0011-9}{%
\path{http://link.springer.com/article/10.1007/s40509-014-0011-9}}.

\bibitem{Russel1} Russell P, Deery O. \newblock \emph{The philosophy of free
will: essential readings from the contemporary debates} \newblock New-York:
Oxford University Press, 2013. \newblock\href{https://books.google.co.uk/books?id=qEVpAgAAQBAJ}%
{\path{https://books.google.co.uk/books?id=qEVpAgAAQBAJ}}.

\bibitem{Georgiev} Georgiev D. \newblock Quantum no-go theorems and
consciousness. \newblock \emph{Axiomathes} 2013; 23 (4): 683--695. \newblock
\href{http://link.springer.com/article/10.1007/s10516-012-9204-1}{%
\path{http://link.springer.com/article/10.1007/s10516-012-9204-1}}.

\bibitem{'tHooft1} 't Hooft G. \newblock The free-will postulate in quantum
mechanics. \newblock arXiv preprint quant-ph/0701097 2007. \newblock \href{http://arxiv.org/abs/quant-ph/0701097}%
{\path{http://arxiv.org/abs/quant-ph/0701097}}.

\bibitem{'tHooft2} 't Hooft G. \newblock The cellular automaton
interpretation of quantum mechanics. A view on the quantum nature of our
universe, compulsory or impossible? \newblock arXiv preprint arXiv:1405.1548
2007. \newblock \href{http://arxiv.org/abs/quant-ph/0701097}{%
\path{http://arxiv.org/abs/1405.1548}}.

\bibitem{Y3} Aharonov Y, Albert DZ, Vaidman L. \newblock How the result of a
measurement of a component of the spin of a spin-1/2 particle can turn out
to be 100. \newblock \emph{Physcal Review Letters} 1988 60(14): 1351--1354. %
\newblock \href{http://dx.doi.org/10.1103/PhysRevLett.60.1351}{%
\path{doi:10.1103/PhysRevLett.60.1351}}.

\bibitem{TC} Tamir B, Cohen E. \newblock Introduction to weak measurements
and weak values. \newblock \emph{Quanta} 2013 2(1): 7-17. \newblock
\href{http://quanta.ws/ojs/index.php/quanta/article/view/14}{%
\path{doi:10.12743/quanta.v2i1.14}}.

\bibitem{Y4} Aharonov Y, Cohen E, Elitzur AC. \newblock Foundations and
applications of weak quantum measurements. \newblock \emph{Physical Review A}
2014 89(5): 052105. \newblock \href{http://dx.doi.org/10.1103/PhysRevA.89.052105}%
{\path{doi:10.1103/PhysRevA.89.052105}}.

\bibitem{Y5} Aharonov Y, Cohen E, Elitzur AC. \newblock Can a future choice
affect a past measurement's outcome? \newblock \emph{Annals of Physics} 2015
355: 258--268. \newblock \href{http://www.sciencedirect.com/science/article/pii/S000349161500055X}%
{\path{doi:10.1016/j.aop.2015.02.020}}.


\bibitem{ECS} Elitzur AC, Cohen E, Shushi T. \newblock The too-late-choice experiment: Bell's proof applied to a time-reversed setting. \newblock Forthcoming. \newblock \href{http://www.ijqf.org/wps/wp-content/uploads/2015/07/Elitzur-IJQF-paper-2.pdf}%
{\path{http://www.ijqf.org/wps/wp-content/uploads/2015/07/Elitzur-IJQF-paper-2.pdf}}.

\bibitem{Gisin} Gisin N, Ribordy G, Tittel W, Zbinden H. \newblock Quantum
cryptography. \newblock \emph{Reviews of Modern Physics} 2002 74(1):
145--195. \newblock\href{http://journals.aps.org/rmp/abstract/10.1103/RevModPhys.74.145}%
{\path{doi:10.1103/RevModPhys.74.145}}

\bibitem{BB84} Bennett C, Brassard G. \newblock Quantum
cryptography: Public key distribution and coin tossing. \newblock in \emph{Proceedings of IEEE International Conference
on Computers, Systems and Signal Processing} 1984 175--179. \newblock\href{http://ci.nii.ac.jp/naid/20001457561}%
{\path{http://ci.nii.ac.jp/naid/20001457561}}



\bibitem{Bousso} Bousso R, Susskind L. \newblock Multiverse interpretation
of quantum mechanics. \newblock \emph{Physical Review D} 2012 85(4): 045007. %
\newblock\href{http://journals.aps.org/prd/abstract/10.1103/PhysRevD.85.045007}{%
\path{doi:10.1103/PhysRevD.85.045007}}.

\bibitem{Deutch} Deutsch D. \newblock The structure of the multiverse. %
\newblock \emph{Proceedings of the Royal Society of London-Series A:
Mathematical, Physical and Engineering Sciences} 2002 458(2028): 2911--2923. %
\newblock\href{http://rspa.royalsocietypublishing.org/content/458/2028/2911.short}{%
\path{doi:10.1098/rspa.2002.1015}}.

\bibitem{Bars1} Bars I. \newblock Survey of two-time physics. \newblock
\emph{Classical Quantum Gravity} 2001 18(16): 3113. \newblock \href{http://iopscience.iop.org/article/10.1088/0264-9381/18/16/303/meta;jsessionid=21E881623B54218A3AFA3B9BC936F425.c1}%
{\path{DOI:10.1088/0264-9381/18/16/303}}.

\bibitem{Bars2} Bars I, Terning J, Nekoogar F, Krauss LM. \newblock
\emph{Extra dimensions in space and time} \newblock pp. 67--87 Heidelberg:
Springer, 2010. \newblock\href{http://link.springer.com/book/10.1007/978-0-387-77638-5}%
{\path{doi:10.1007/978-0-387-77638-5}}.


\bibitem{Carlini} Carlini A, Frolov VP, Mensky MB, Novikov ID, Soleng HH. %
\newblock Time machines: the principle of self-consistency as a consequence
of the principle of minimal action. \newblock \emph{International Journal of
Modern Physics D} 2012 4(05): 557--580. \newblock\href{http://www.worldscientific.com/doi/abs/10.1142/S0218271895000399}%
{\path{doi:10.1142/S0218271895000399}}.


\bibitem{Novikov} Novikov ID. \newblock Time machine and self-consistent
evolution in problems with self-interaction. \newblock \emph{Physical Review
D} 1992 45(6): 1989-1994. \newblock\href{http://journals.aps.org/prd/abstract/10.1103/PhysRevD.45.1989}%
{\path{doi:10.1103/PhysRevD.45.1989}}.

\bibitem{Lloyd} Lloyd S, Maccone L, Garcia-Patron R, Giovannetti V, Shikano Y. \newblock Quantum mechanics of time travel through post-selected teleportation. \newblock \emph{Physical Review D} 2011 84(2): 025007. \newblock\href{http://journals.aps.org/prd/abstract/10.1103/PhysRevD.84.025007}%
{\path{doi:10.1103/PhysRevD.84.025007}}.

\bibitem{Y6} Aharonov Y, Popescu S, Tollaksen JM. \newblock Each instant of
time a new universe. \newblock In \emph{Quantum Theory: A Two-Time Success
Story}, Struppa DC, Tollaksen JM (editors), pp. 21--36. Milan: Springer,
2014. \newblock \href{http://link.springer.com/chapter/10.1007/978-88-470-5217-8_3}%
{\path{10.1007/978-88-470-5217-8_3}}.

\end{thebibliography}
\end{document}